\documentclass[10pt,letterpaper,iop]{emulateapj}

\shorttitle{Kappa Fornaci, a triple radio-star}
\shortauthors{Tokovinin}

\begin{document}

\title{Kappa  Fornaci,  a  triple radio-star \footnotemark[1]}

\footnotetext[1]{Based  on
    observations  obtained  with  CHIRON  spectrometer at the 1.5-m  CTIO
    telescope  operated by  SMARTS (NOAO  program 2012B-0075),  at the
    SOAR   telescope,   and  at   the   Gemini  Observatory   (program
    GS-2012B-Q-71, PI M.~Hartung)}

\author{Andrei Tokovinin}
\affil{Cerro Tololo Inter-American Observatory, Casilla 603, La Serena, Chile}
\email{atokovinin@ctio.noao.edu}

\begin{abstract}
Bright and  nearby (22\,pc) solar-type  dwarf $\kappa$~For (HIP~11072)
is  a triple  system. The  close pair  of M-type  dwarfs Ba,Bb  with a
tentative period  of 3.7 days moves  around the main component  A on a
26-year orbit. The mass of  the ``dark companion'' Ba+Bb is comparable
to the  mass of A,  causing  large  motion of the  photo-center.  The
combined  spectro-interferometric  orbit  of  AB is  derived  and  the
relative  photometry of  the  components A  and  B is  given.  A  weak
signature of Ba  and Bb is detected in  the high-resolution spectra by
cross-correlation  and by  variable emission  in the  Bahlmer hydrogen
lines.   The activity  of  the M-dwarfs,  manifested  by a  previously
detected  radio-flare, is  likely maintained  by  synchronisation with
their tight orbit.  We discuss  the frequency of similar hidden triple
systems,  methods  of  their   detection,  and  the  implications  for
multiple-star statistics.
\end{abstract}


\section{Introduction}
\label{sec:intro}

\setcounter{footnote}{1}

In January  1993 \citet{Guedel} detected microwave  radiation from the
nearby  solar-type   dwarf  $\kappa$~For.   Its   source  was  located
$0\farcs23$ south of the star.  At that time it was already known that
$\kappa$~For  is a  spectroscopic binary.   So, the  authors suggested
that the radio emission comes from the secondary companion, presumably
a low-mass flaring dwarf. As we show below, the secondary is in fact a
tight pair  of M dwarfs.   Despite the old  age of this  triple system
(4 to 6~Gyr), fast axial rotation of the M-dwarfs (hence high
activity) is maintained by synchronization with the orbit.

Kappa     Fornaci     (HIP~11072,     HD~14802,     HR~895,     GJ~97,
$\alpha_{2000}=$2:22:32.54, $\delta_{2000}=-$23:48:58.8)  is located at
a  distance of  22\,pc from  the Sun  according to  the  original {\it
  Hipparcos}   catalog  \citep{HIP}  and   its  new   reduction
\citep{HIP2}.   The   {\it  Hipparcos}  satellite   detected  a  large
acceleration  of 19.4\,mas~yr$^{-2}$ presumably  caused by  the invisible
(astrometric)    companion    \citep{MK05}.     \citet{Gontcharov2001}
collected  historical astrometric  data  spanning half  a century  and
suggested  that   the  orbital  period  is   around  26.5\,yr.   Later
\citet{GK02} published the elements of the astrometric orbit with this
period  and with  a  relatively  large semi-major  axis  of $\alpha  =
0\farcs26$, noting that the ``dark companion'' is massive.

Radial velocity (RV) of $\kappa$~For was monitored both at Lick and at
La Silla in search of planetary companions. None was found so far, but
the   RV  trend  caused   by  the   stellar  companion   was  obvious.
\citet{Nidever02}  noted the trend,  while \citet{Endl02}  published a
preliminary spectroscopic  orbit with a  21\,yr period.  \citet{Abt06}
added their own observations  and revised the orbit slightly, although
the  full  orbital  cycle  was  not yet  covered.   Neither  of  these
publications mentions the astrometric results.

The  astrometric and  spectroscopic  companion was  first resolved  by
\citet{LAF2007} in  2005 and, independently, by  \citet{TC08} in 2007.
For this reason it  received two confusing ``discoverer'' codes LAF~27
and    TOK~40    in     the    Washington    Double    Star    Catalog
\citep{WDS}.\footnote{The  WDS keeps  outdated tradition  of assigning
  discoverer  codes even  when the  duplicity was  actually discovered
  before  the   first  position  measurement,   as  in  the   case  of
  $\kappa$~For} By using all  resolved measures and fixing the orbital
period  and   eccentricity  to   those  of  the   Gontcharov's  orbit,
\citet{HTM12} computed the first visual orbit of grade 4.  Here we add
new observations and  derive the combined interferometric/spectroscopic
orbit which  agrees with the  astrometric orbit.

SIMBAD lists  187 references  to date and  provides basic data  on the
primary  companion, such  as spectral  type G1V,  magnitudes $V=5.19$,
$B-V=0.60$, $K_s=3.741$, and near-solar metallicity 
\citep[see  the  full compilation  of  stellar parameter  measurements
  in][]{PASTEL}. The  star is located  above the Main Sequence  in the
$(V,  B-V)$ color-magnitude  diagram (CMD),  being  obviously evolved.
\citet{N04}  estimate   the  age  of  6\,Gyr   by  isochrone  fitting.
\citet{Nielsen10} cite  indirect age estimates of  4\,Gyr and 6.3\,Gyr
from the  lithium line strength and rotation,  respectively.  The star
was  observed  by {\it  Spitzer}  and found  to  have  no debris  disk
\citep{Trilling08}.   Yet, \citet{Nakajima2012} claimed  recently that
$\kappa$~For is  young and  belongs to the  IC~2391 moving  group.  In
calculating the kinematic parameters they overlooked the companion and
used the {\it Hipparcos} proper  motion (PM), biased by the orbit.  We
calculate   the   heliocentric   Galactic   velocity   components   of
$\kappa$~For from the center-of-mass PM \citep{Gontcharov2001} and the
new    $\gamma$-velocity   to   be    $[U,V,W]   =    [-19.5,   -16.2,
  -9.6]$\,km~s$^{-1}$ ($U$ is positive towards the Galactic center).

\begin{figure*}
\plotone{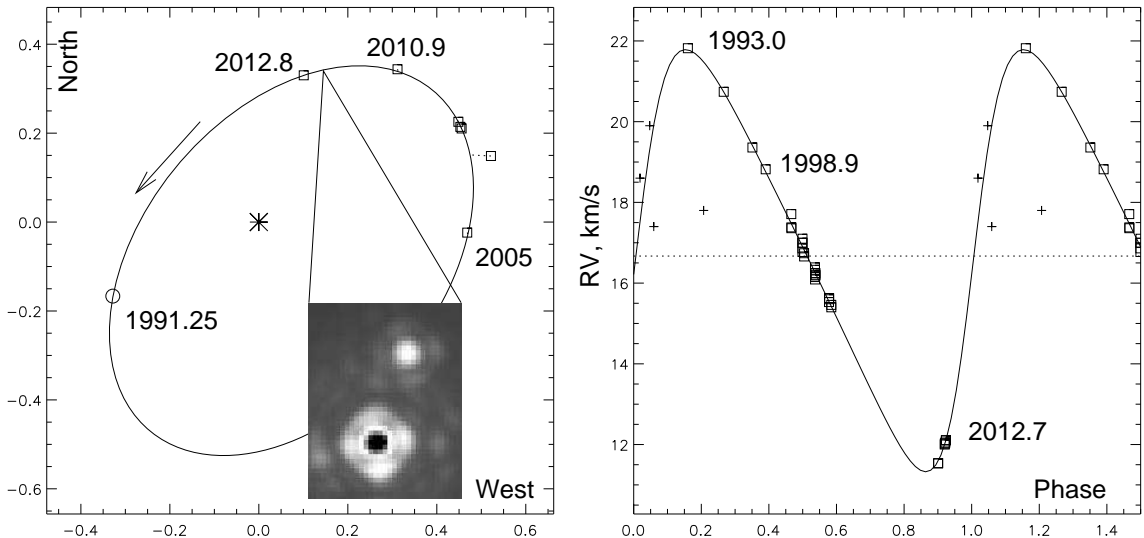}
\caption{Combined orbit of $\kappa$~For  AB.  Left panel: orbit in the
  plane of the sky, scale in arcseconds. The insert shows the $K$-band
  image taken at Gemini-S on  2012.67.  Right panel: the RV curve with
  early measures plotted as crosses.
\label{fig:orbit} 
}
\end{figure*}

In Section~2 we derive the updated  orbit of the outer system AB. Then
in Section~3  we evaluate  the parameters of  the companions  and show
that  the   secondary  B  is   over-massive,  being  itself   a  close
binary.  This   hypothesis  is   confirmed  in  Section~4   by  direct
spectroscopic detection of Ba and  Bb. In the last Section~5
we  discuss the  implications  of this  finding  for   multiplicity
statistics and  exo-planet searches.

\section{The orbit of AB}
\label{sec:orb}

So far,  spectroscopic, astrometric, and visual  orbits of $\kappa$~For
were  determined almost  independently of  each other.   We  add fresh
measures  and combine available  data in  the single  orbital solution
presented  in Fig.~\ref{fig:orbit}.   The orbital  elements  and their
formal errors are obtained by the unconstrained least-squares fit with
weights inversely proportional to the measurement errors. The elements
are  listed  in Table~\ref{tab:orbits}  together  with the  previously
published  orbits.    We  have  not  used  the   astrometric  data  of
\cite{GK02},  so the  good match  between  spectro-interferometric and
astrometric orbits speaks to their  veracity.  This is not the case of
spectroscopic  orbits which  have a  too short  period.  \citet{Abt06}
mention that they used ``some older measures'' in the orbit derivation
but do  not specify their source.   In the future, with  better RV and
positional coverage, the elements will  be determined to a much higher
precision than  here, but  they will no  longer be revised  beyond the
stated  errors.  The  new orbit  and the  {\it Hipparcos}  parallax of
45.53$\pm$0.82\,mas \citep{HIP2} correspond  to the semi-major axis of
11.44\,AU  and  to  the  mass  sum  of  2.25$\pm$0.13\,$M_\odot$  (the
uncertainty comes mostly from the parallax error).

\begin{deluxetable*}{ ccc ccc ccc l }
\tabletypesize{\scriptsize} 
\tablecaption{Orbits of $\kappa$~For AB
\label{tab:orbits}}
\tablehead{
$P$ & $T$ & $e$ & $a$ &  $\Omega_A$ & $\omega_A$ & $i$ & $K_1$ & $\gamma$ &
  Ref. \\
(yr) & (BY) &  & ($''$) & $^\circ$ & $^\circ$ & $^\circ$ & km~s$^{-1}$ & km~s$^{-1}$
  & }
\startdata
26.5$\pm$2 & 1983.6$\pm$2.3 & 0.2$\pm$0.1 & 0.26$\pm$0.01 & 133$\pm$6 & 202$\pm$31 & 56$\pm$4 & \ldots & \ldots & a \\
 21.46$\pm$0.09 & 1967.26$\pm$1.9 &  0.30$\pm$0.04 & \ldots & \ldots & 263$\pm$12 & \ldots & 3.95$\pm$0.26 & 18.13$\pm$0.66 & b \\
26.5 * & 2015.69 &  0.20 * &  0.493 &  131.6 &  275.5  & 44.8 & \ldots & \ldots & c \\
25.81$\pm$0.15 &1988.89$\pm$0.17 & 0.339$\pm$0.013 & 0.521$\pm$0.004 &
139.8$\pm$1.4 & 266.3$\pm$1.0 & 50.4$\pm$0.5 & 5.23$\pm$0.13 &
16.67$\pm$0.06 & d 
\enddata
\tablenotetext{}{References: a -- astrometric \citep{GK02}; b -- spectroscopic
  \citep{Abt06}; c -- visual \citep{HTM12}; d -- combined, this work}
\end{deluxetable*}

\begin{deluxetable}{l ccc l}
\tabletypesize{\scriptsize} 
\tablewidth{0pt}
\tablecaption{Radial velocities  and residuals of $\kappa$~For A
\label{tab:RV}}

\tablehead{JD & RV & $\sigma$ & O$-$C & Ref. \\
+2400000 & km~s$^{-1}$ & km~s$^{-1}$ & km~s$^{-1}$ & }  
\startdata
19378.67  &    18.6  & 2.0  & 1.08  & a \\
19649.90  &    19.9  & 2.0  & 0.58  & a \\
19762.62  &    17.4  & 2.0  & -2.52 & a \\
21152.78  &    17.8  & 10.0 & -3.69 & a \\
 49000.0  &    21.82 &  0.2 & 0.04  & b \\
 50000.0  &    20.74 &  0.2 & 0.00  & b \\
 50800.0  &    19.36 &  0.2 & -0.05 & b \\
51170.080 &   18.824 &  0.1 & 0.06  & c \\
51884.735 &  17.71   & 0.1  & 0.23  & d \\
\ldots    &   \ldots & \ldots & \ldots & d  \\         
53001.665 &   15.46  & 0.1  & 0.04  & d \\
55979.533 &   11.53  & 0.5  & -0.055 & e \\
55983.528 &   11.54  & 0.5  & -0.052 & e \\
56165.830 &   12.001 & 0.01 & -0.005 & e \\
56167.865 &   12.019 & 0.01 &  0.007 & e \\
56171.797 &   12.032 & 0.01 &  0.008 & e \\
56182.861 &   12.048 & 0.01 & -0.009 & e \\
56194.830 &   12.095 & 0.01 &  0.002 & e \\
56200.834 &   12.111 & 0.01 & -0.001 & e 
\enddata
\tablenotetext{}{References: a --  \citep{Lick}; b -- \citep{Endl02}, zero-point 20.36\,km~s$^{-1}$;
  c -- \citep{Nidever02}; d -- \citep{Abt06}; e -- this work.}
\end{deluxetable}

\begin{deluxetable}{l rrr   c c l}
\tabletypesize{\scriptsize} 
\tablewidth{0pt}
\tablecaption{Positional measures and residuals of $\kappa$~For AB
\label{tab:speckle}}

\tablehead{$t$ & $\theta$ &  $\rho$  & $\sigma$ &
  \multicolumn{2}{c}{(O$-$C)$_\theta, \rho$}  & Ref. \\ 
yr  &  $^\circ$ & mas & mas &   $^\circ$ & mas &   }
\startdata
2005.6348 & 267.1 & 469  & 5 &     0.4    & -2    & a\\
2007.8130 & 285.9 & 542 & 100 &  -1.7    &  44   &  b \\
2008.6282 & 294.8 & 502  & 2 &    -0.2    &  0    & c\\
2008.6936 & 295.3 & 501  & 2 &    -0.3    & -1    & c\\
2008.8549 & 296.7 & 502  & 2 &    -0.4    &  0    & c\\
2010.9655 & 317.8 & 464  & 2 &     0.6    &  3    & d\\
2012.8300 & 343.0 & 345  & 2 &     0.1    &  0    & e 
\enddata
\tablenotetext{}{References: a -- \citep{LAF2007}; b -- \citep{TC08};
  c -- \citep{TMH10}; d -- \citep{HTM12}; e -- this work.}
\end{deluxetable}

Radial  velocities  used  in  the  orbit  calculation  are  listed  in
Table~\ref{tab:RV}.   \citet{Endl02} did  not publish  the  RVs; three
points from  their Fig.~6  (1993--1998) are nevertheless  included in
the present solution by assuming  that the zero velocity in their plot
corresponds to  20.36\,km~s$^{-1}$. In  this case their  data match  well the
single  measurement in  1999  published by  \citet{Nidever02} and  the
velocities from  \citep{Abt06}, of which  we list here only  the first
and the last points. The latest RV measures are obtained by the author
in 2012 (see Section~\ref{sec:chiron}).   We also use the RVs measures
made  in   Santiago  in  1911--1916  with   the  2-prism  spectrometer
\citep{Lick}.  Although the precision  of these early data (crosses in
Fig.~\ref{fig:orbit},  right) is  low, they help in constraining the orbital period.

Positional measures of $\kappa$~For AB  and their residuals to the new
orbit  are listed  in  Table~\ref{tab:speckle}. The  first measure  is
performed with  adaptive optics, the remaining data  come from speckle
interferometry at  the SOAR telescope  in Chile.
 The latest speckle measure was made at SOAR on October
29,  2012.   The first  speckle  resolution  is  given a  low  weight,
considering uncertain calibration in that early experimental work.

\section{Properties of the  components}
\label{sec:Bab}

The component  A ($V=5.19$~mag, $B-V=0.60$) is located  $\sim 1.5^m$ above
the Main Sequence in the $(M_V, B-V)$ CMD, as noted by \citet{N04}.  By
fitting isochrones, these authors estimated  the mass of the main star
$M_{\rm A}$ to  be between 1.12 and 1.18\,$M_\odot$.   We checked this
against the isochrones of \citet{Girardi} and found that the mass of A
can be  as high as  1.25\,$M_\odot$ if its  age is 4\,Gyr (it  is just
leaving the  Main Sequence).  It is  safe to assume that  $M_{\rm A} =
1.20 \pm 0.05\,M_\odot$.

\begin{deluxetable}{l | c c c c}
\tabletypesize{\scriptsize} 
\tablecaption{Photometry of the resolved components
\label{tab:ptm} }

\tablehead{Parameter & $V$ & $I$ & $H$ & $K_s$ \\
                     & mag & mag & mag   & mag }
\startdata
$m_{\rm A+B}$          & 5.19 & 4.51 & 3.712 & 3.741 \\ 
$m_{\rm B} - m_{\rm A}$ & 5.02$\pm$0.04 & 3.69$\pm$0.05 & (2.46) & (2.14) \\
$m_{\rm B}$            & (10.21)         & (8.23) & 6.3$\pm$0.2 & 6.0$\pm$0.2   
\enddata
\end{deluxetable}

The  photometric  data  are  collected  in  Table~\ref{tab:ptm}.   The
magnitude difference  in the Str\"omgren  $y$ and Cousins  $I$ filters
was determined by speckle interferometry  at SOAR (6 and 5 independent
measures  respectively); we  list here  the average  values  and their
formal  errors and  assume $\Delta  V \approx  \Delta y$.  The derived
quantities   are  listed  in   brackets.   

The flux ratio  in the 1.54-1.65\,$\mu$m band (which  is close to $H$)
was estimated by \citet{LAF2007} as $\Delta  H = 2.7 \pm 0.2$~mag. On 2012
September 2 (2012.6706) the binary was resolved with the NICI adaptive
optics  instrument \citep{Chun08},  see Fig.~\ref{fig:orbit}.   As the
primary  companion  was heavily  saturated,  the position  measurement
($340.5^\circ$, 0\farcs342)  is not very accurate (it  is not included
in Table~\ref{tab:speckle}), and no relative photometry could be made.
However,  the flux  from  the well-resolved  companion  B (after  halo
subtraction)  was estimated  by  comparing it  to  stars HIP~8674  and
HIP~12425 observed  before and  after this target  at nearly  the same
airmass. The  $H$ and $K_s$ magnitudes  of those two  stars from 2MASS
were used  to determine the zero  point (the actual  wavelength of the
narrow-band filters was 1.587\,$\mu$m and 2.272\,$\mu$m).  In this way
we  obtained crude  estimates of  the companion's  infrared magnitudes
$H=6.0\pm0.2$~mag and $K_s = 6.3\pm0.2$~mag.

The  magnitudes and  colors of  the companion  B are  thus established
reasonably well, $(V - I)_{\rm B} = 1.98 \pm 0.06$ and $(V - K_s)_{\rm
  B} = 4.2 \pm 0.2$. Standard relations match those colors for a dwarf
star of  $\sim$0.48\,$M_\odot$, but the  component B is  located about
$1.5^m$ above the  Main Sequence in the $(M_V,  V-I)$ and $(M_V, V-K)$
CMDs;   its   luminosity   corresponds    to   a   single   dwarf   of
$\sim$0.65\,$M_\odot$.   This  discrepancy  is  caused by  the  binary
nature of B.

Large  total  mass  of  the  companion  B  follows  from  the  orbital
parameters.   Motion  of  the  photo-center is  characterized  by  the
semi-major axis $\alpha$  of the astrometric orbit.  Its  ratio to the
semi-major axis $a$  of the visual orbit is related  to the mass ratio
$q = M_{\rm B}/M_{\rm A}$ and to the light ratio $r$ as
\begin{equation}
\phi = \frac{\alpha}{a} = \frac{q - r}{(q+1)(r+1)}.
\label{eq:phi}
\end{equation}
In our  case we can neglect  the companion's light  because $r \approx
0.01$.   The orbit of  AB derived  here and  the astrometric  orbit of
\citet{GK02} correspond  to $\phi = 0.50$  and $q = \phi/(1  - \phi) =
1.0$.   The companion  B  is therefore  as  massive as  A while  being
$\sim$100 times  fainter at optical wavelengths.   This finding agrees
with the large orbital mass sum.

Another argument  for the high  companion's mass is furnished  by radial
velocities.   Using the  formula $a_1  \sin  i =  0.01375 K_1  P (1  -
e^2)^{0.5}$ ($a$  in $10^6$km, $K_1$  in km~s$^{-1}$, $P$ in  days) we
obtain $a_1=  5.53$\,AU and $\phi=a_1/a  = 0.48$, leading  to $q=0.93$.

The {\em Hipparcos} astrometry agrees  very well with the new orbit if
we adopt the companion's  masses inferred from the photometry, $M_{\rm
  A}= 1.20\,M_\odot$  and $M_{\rm B}= 0.96\,M_\odot$,  or $q=0.80$ and
$\phi = 0.43$.  Using this  $\phi$, we compute the photo-center motion
during  the  {\it Hipparcos}  mission  (mean  epoch 1991.25,  duration
3.2\,yr). Table~\ref{tab:hip} shows that  the orbital proper motion of
the photo-center  $\Delta \mu$ and its  acceleration $\dot{\mu}$ match
the  ephemeris  in  both  magnitude  and direction.   In  1991.25  the
companion   moved   mostly  to   the   South   (see   the  circle   in
Fig.~\ref{fig:orbit}, left), the photo-center moved to the North.  The
measured $\Delta  \mu$ is the  difference between the  {\it Hipparcos}
PM of $(+197,  -5)$\,mas~yr$^{-1}$ and the long-term PM
of $(+196, -60)$\,mas~yr$^{-1}$ derived by \citet{Gontcharov2001} from
the  combination  of  all  ground-based  data.   The  {\it  Hipparcos}
astrometry thus confirms the orbit and the fact that B is massive.

\begin{deluxetable}{l | c c | c c}
\tabletypesize{\scriptsize} 
\tablecaption{Hipparcos astrometry and the orbit
\label{tab:hip} }

\tablehead{Parameter & \multicolumn{2}{c|}{ \it Hipparcos}  &
  \multicolumn{2}{c}{Orbit ($\phi = 0.43$)} \\
                     & RA & Dec & RA & Dec  }
\startdata
$\Delta \mu$, mas~yr$^{-1}$ & 1 & +55 & $-$9 & +56 \\
$\dot{\mu}$, mas~yr$^{-2}$  & +13.0$\pm$2.0  & $-$14.4$\pm$1.8 & +17.7 & $-$9.2   
\enddata
\end{deluxetable}

\section{Spectroscopy}
\label{sec:chiron}

\subsection{Observations}

Knowing  that the  companion B  must be  a close  binary, we  tried to
detect   this   sub-system   spectroscopically,  despite   its   small
contribution to  the combined  light. Optical spectra  with resolution
$\lambda/\Delta \lambda  = 80\,000$  were recorded with  the fiber-fed
CHIRON echelle  spectrometer installed at the 1.5-m  telescope at CTIO
\citep{CHIRON} and operated in  service mode.  The object was observed
in  2012 six times,  on August  25, 27,  31 and  September 11,  23, 29
(hereafter nights  1 to 6).  On  each visit, two  300-s exposures with
the image slicer were taken, accompanied by the comparison spectrum of
the thorium-argon  (ThAr) lamp.  During  this period, the  position of
the ThAr spectrum  remained stable on the CCD to  better than 1 pixel.
Moreover, on the  first night the star HIP~14086 was  observed as a RV
reference.

Extracted  and  wavelength-calibrated spectra  were  delivered by  the
pipeline  running at  the Yale  University.  They  contain  59 echelle
orders (central wavelengths from  4605\AA~ to 8713\AA) of 3200 pixels
each (1.0768\,km~s$^{-1}$ per pixel).  As red orders with $\lambda >7000$\AA~
suffer  from  extraction and  calibration  problems  (too few  thorium
lines), they are not used  here.  The maximum intensity in the spectra
varies from night to night with a total range of two times (from 25 to
50 thousand electrons per pixel).   This corresponds to a S/N from 160
to 220.

\subsection{Radial velocities}

\begin{figure}[ht]
\epsscale{1.0}
\plotone{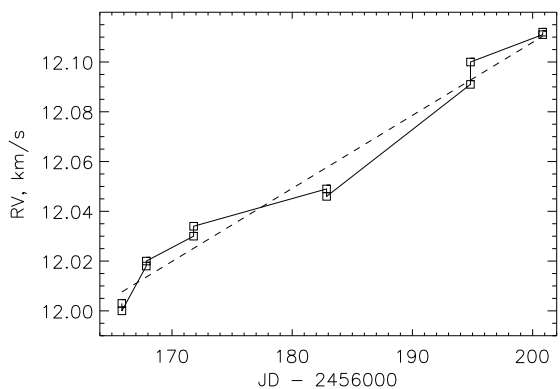}
\caption{ Radial  velocity of  the component A  vs. time.   The dashed
  line shows a linear fit with a slope of 2.9\,m~s$^{-1}$ per day.
 \label{fig:RV}}
\end{figure}

Radial  velocities of $\kappa$~For  were derived  by cross-correlating
the spectra with  a binary mask which equals one in  the lines of the
solar spectrum and  zero otherwise. The mask was  constructed from the
digital solar spectrum  \citep{Arcturus}\footnote{NOAO data archive:
  \url{ftp://ftp.noao.edu/catalogs/arcturusatlas}},  binary-clipped at
0.6 of the continuum level.   The cross-correlation is computed in the
wavelength range  from 4600\AA ~to  6500\AA, thus avoiding  the 
contamination by telluric lines  at longer wavelengths. Regions within
$\pm  0.5$\AA~ of  the hydrogen  Bahlmer lines  are excluded  from the
mask.   The  minimum  in   the  cross-correlation  function  (CCF)  is
approximated  by a  Gaussian  curve, its  center  is taken  to be  the
apparent  stellar radial  velocity,  which is  then  corrected to  the
barycenter of  the solar system  in the standard way.   This procedure
relies  on the  wavelength  calibration of  the  reduced spectra.   We
attempted to  refine the velocity zero point  by cross-correlating red
orders  with  the mask  of  telluric lines,  but  did  not obtain  any
trustworthy   results,  probably  because   of  the   poor  wavelength
calibration in the red.

Figure~\ref{fig:RV} shows  the RVs derived from  12 individual spectra
as a  function of  time (the average  RV for  each night is  listed in
Table~\ref{tab:RV}). The linear trend of 2.9\,m~s$^{-1}$ per day is caused by the
orbital motion.  The  rms scatter of RV around  this line is 6.8\,m~s$^{-1}$,
the rms residual  to the orbit is 6.7\,m~s$^{-1}$.  As  measures from the two
nightly spectra agree  well, most of the scatter  can be attributed to
the wavelength  calibration based on the  ThAr lamp.  The  mean FWHM of
the Gaussian fits to the CCF is 12.96\,km~s$^{-1}$, their equivalent width is
5.59\,km~s$^{-1}$.

The same procedure applied to HIP~14086 gives an RV of $+39.14$\,km~s$^{-1}$.
According  to \citet{Nidever02}  the RV  of this  star is  constant at
$+42.718$\,km~s$^{-1}$,  but  \citet{N04} list  a  quite  different value  of
$+38.0$\,km~s$^{-1}$ (while SIMBAD lists the wrong RV of $-38.4$\,km~s$^{-1}$).  Apparently,
this star is a  yet unrecognized slow spectroscopic binary, unsuitable
as a  RV standard.  However,  we use our  RV measure of  HIP~14086 to
derive the RV  of $\kappa$~For from the two  spectra taken in February
2012 during CHIRON tests  (those spectra lack wavelength calibration).
These two  measures with a somewhat  uncertain zero point are  given low
weight in the orbit solution by adopting  errors of 0.5\,km~s$^{-1}$.

\subsection{Variability of the hydrogen line profile}

We  detect a  tiny  variability of  the  H$\alpha$ profile  presumably
caused  by the  moving emission  of  Ba and  Bb. This  is achieved  by
subtracting the  template obtained by averaging  all spectra together.
Although the  spectrometer was very stable  during these observations,
the  spectrum moved  on  the CCD  by  12\,km~s$^{-1}$ owing  to the  variable
barycentric correction.  The template  was constructed by shifting the
nightly-averaged  spectra (with  a  3-pixel median  smoothing of  each
spectrum to remove the cosmic-rays  spikes)  to match the night 1,
normalizing them  in intensity,  and adding together.   Obviously, the
telluric  lines  do not  move  together  with  the stellar  lines  and
therefore show up in the  residuals between the individual spectra and
the template.  The rms residuals after template  subtraction are 0.7\%
in  the order 37  containing H$\alpha$  and 0.5\%  in the  blue orders
which are free from the telluric lines.

\begin{figure}[ht]
\epsscale{1.0}
\plotone{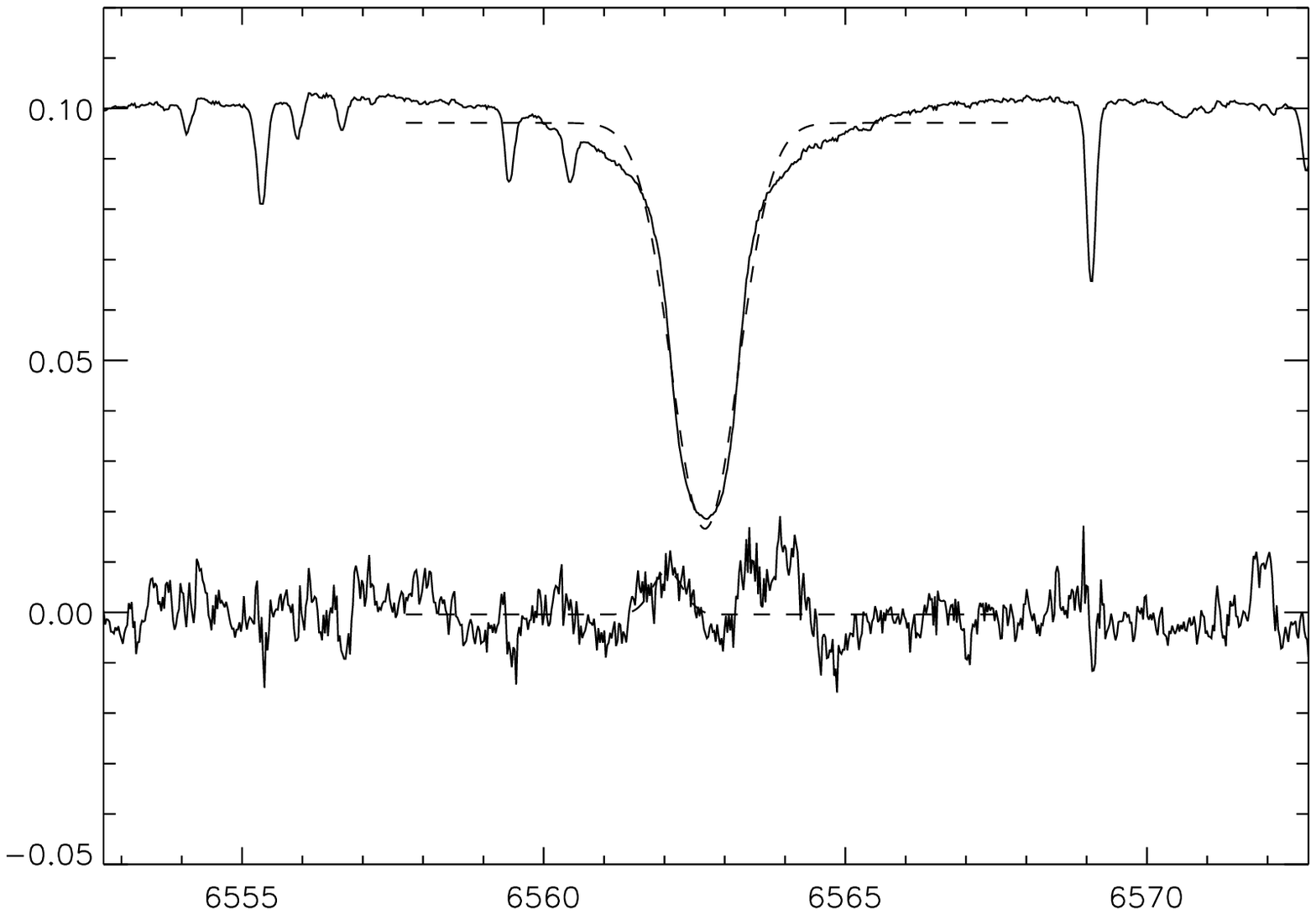}
\plotone{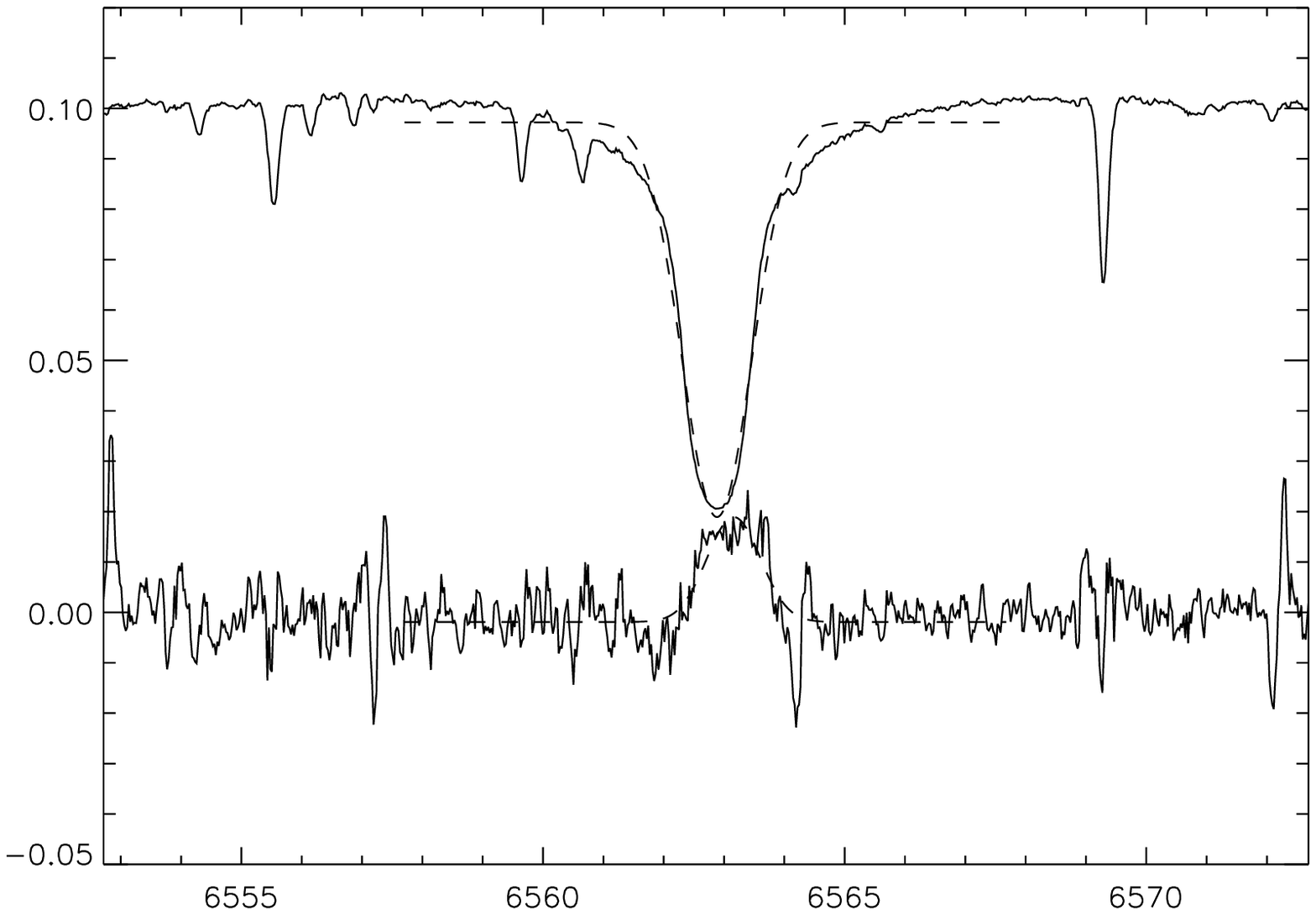}
\caption{Variable features in  the H$\alpha$ line on the  night 1 (top
  panel) and on  the night 5 (bottom panel).  In  each panel the upper
  curves show the continuum-normalized  spectrum scaled by 0.1, with a
  Gaussian fit to  the H$\alpha$ (dashed line). The  lower curves show
  the residuals after template subtraction.
 \label{fig:Ha}}
\end{figure}

We see a  clear residual signal in the H$\alpha$ line  on the nights 1
and 5, while  on the remaining nights the residuals  are close to zero
(Fig.~\ref{fig:Ha}). Similar features of lower amplitude are also seen
in  the   H$\beta$  line,   confirming  their  reality.    The  single
``emission''  peak on  the night  5  can be  approximated by  Gaussian
curves. Their amplitude is 2.1\%  and 1.3\% for H$\alpha$ and H$\beta$
respectively, and they are shifted relative to the stellar spectrum by
$+0.28$\AA~ and $+0.23$\AA~  which corresponds to $+12.8$\,km~s$^{-1}$
and   $+14.2$\,km~s$^{-1}$.   If  the   emission  originated   on  the
slowly-rotating primary companion A  ($V \sin i = 4$\,km~s$^{-1}$), it
would be  centered on the stellar  line rather than  shifted.  We show
below that  on the night  5 the lines  of Ba and Bb  were superimposed
near  their  center-of-mass  velocity  which should  be  displaced  by
$+9.2$\,km~s$^{-1}$ relative  to the  RV of A  according to  the orbit
(the masses of A and Ba+Bb  being near-equal).  The double peak on the
night 1 also matches the expected position of the emission from Ba and
Bb. The  non-detection of emission  features on the other  four nights
could possibly be explained by variable activity of the M-dwarfs.

\subsection{Retrieval of the Ba and Bb signature by cross-correlation}

\begin{figure}[ht]
\epsscale{1.0}
\plotone{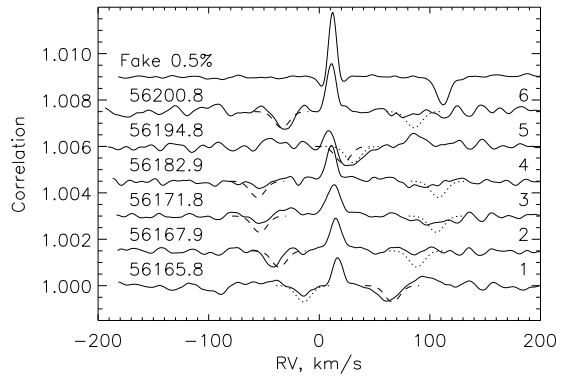}
\caption{Correlation functions of residuals with the Arcturus mask for
  6  nights (from  bottom up,  displaced  by 0.0015  per night).   The
  horizontal axis corresponds to the heliocentric RV.  The upper curve
  is correlation with the template  to which a fake companion of 0.5\%
  intensity and  +100\,km~s$^{-1}$ RV shift is added.   The dashed and
  dotted lines  show suggested positions of the  dips corresponding to
  Ba and Bb components respectively  moving on a 3.7-d orbit.  Numbers
  on the left are Julian dates, numbers on the right are the nights.
 \label{fig:res1}}
\end{figure}

We detect the signature of absorption lines belonging to  Ba and
Bb  by  correlating  the   residual  spectra  (after  subtracting  the
template, see above)  with the solar or Arcturus  masks (the latter is
analogous  to the  solar binary  mask  but uses  the digital  Arcturus
spectrum from the same source). Only wavelengths shorter than 6500\AA~
are  used in the  correlation.  The  resulting CCF  has a  weak narrow
feature centered on  the RV of A (Fig.~\ref{fig:res1}).  This
feature originates because the  resolution of the template spectrum is
slightly  less than  the resolution  of the  nightly spectra  owing to
small residual  alignment errors in  the template creation.   The sign
and  intensity  of this  feature  depend  on  the relative  degree  of
smoothing applied to the spectra and the template.

Apart from the  central feature, we see additional  dips with variable
position and  intensity, presumably produced  by Ba and  Bb.  Gaussian
approximations  of  the  strongest  dips  have  FWHM  between  15  and
23\,km~s$^{-1}$  and equivalent  width  of about  0.0017\,km~s$^{-1}$.
The star  gives a CCF  with equivalent width of  3.54\,km~s$^{-1}$ for
the Arcturus mask, so the dip ratio corresponds to the flux difference
on the  order of $5.8^m$ (we  neglect here the different  match of the
Arcturus mask with  the late-type spectrum of Ba on  one hand and with
the G0V spectrum  of A on the other hand).  A  fake companion of 0.5\%
relative  intensity ($5.75^m$  fainter  than the  primary) is  readily
detectable    by   this    method    (see   the    upper   curve    in
Fig.~\ref{fig:res1}).

The  curves in  Fig.~\ref{fig:res1}  give an  impression of  secondary
lines moving  with an amplitude  of $\sim$80\,km~s$^{-1}$. We  found a
tentative  circular  orbit  of   Ba,Bb  with  a  period  of  3.666\,d,
semi-amplitude   $K_1 =  K_2 =   83$\,km~s$^{-1}$,   $\gamma$-velocity  
+26\,km~s$^{-1}$,  and initial  epoch (RV  maximum)  JD~2456166.45.
The data  at hand are not  sufficient for the  orbit determination and
the above elements  are a guess only.  The dashed  and dotted lines in
Fig.~\ref{fig:res1} are  Gaussian dips corresponding to the  Ba and Bb
components  respectively and  positioned according  to  this tentative
orbit  (not  fitted to  the  CCF).   The  assumed amplitudes  of  both
Gaussians are $-$0.0007, their  FWHM is 16\,km~s$^{-1}$.  On the night
1  we clearly  see  both dips.  On  the nights  3 and  4  the CCFs  of
residuals  look  almost  identical,  the  dips are  near  the  maximum
separation.
Remember that the template includes the average spectrum of Ba and Bb,
so their  lines near maximum separation (the  most frequent situation)
are partially subtracted, leaving  only small residuals.  On the night
5 the dips overlap near the $\gamma$-velocity.

\begin{figure}[ht]
\epsscale{1.0}
\plotone{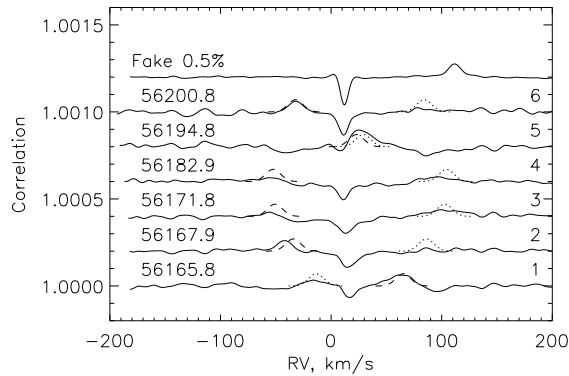}
\caption{Correlation functions of  residuals with the M-star synthetic
  spectrum   for    6   nights   (from   bottom    up),   similar   to
  Fig.~\ref{fig:res1}.
 \label{fig:res3}}
\end{figure}

The reality of  the detection of Ba and Bb  was checked by correlating
the residuals  with the synthetic  spectrum of a late-type  dwarf.  We
used          the          $R=500\,000$         spectrum          from
\citep{Bertone08}\footnote{\url{http://www.inaoep.mx/\~{}modelos/bluered/documentation.html}}
with  parameters  $T_e   =  4000$\,K,  $\log  g  =   5.0$,  and  solar
metallicity.  The  resulting CCF should  have a peak  corresponding to
the velocity of the  secondary.  Indeed, we see in Fig.~\ref{fig:res3}
positive  features at  approximately same  velocities as  the  dips in
Fig.~\ref{fig:res1}.  To  highlight these features,  we again over-plot
fiducial  Gaussians  with  an  amplitude of  +0.00007  and  velocities
predicted by the tentative Ba,Bb orbit.

The case of  the component B being a binary is  very strong. Its high
mass, position  in the CMD,  variable emission, and  moving absorption
lines  all  confirm  this  hypothesis.   The RV  semi-amplitude  in  a
spectroscopic binary with a circular  edge-on orbit of period $P$ days
is $A_0 = 213 P^{-1/3}  M_2 (M_1 + M_2)^{-2/3}$\,km~s$^{-1}$, or 69\,km~s$^{-1}$ for
$P=3.7$\,d, assuming  $M_1 = M_2 =  0.5 M_\odot$. The  RV amplitude of
our tentative  Ba,Bb orbit is therefore in  qualitative agreement with
its period.

\section{Discussion}
\label{sec:disc}

The triple  system $\kappa$~For consists of the  main slightly evolved
(age 4--6\,Gyr) component A of  1.2\,$M_\odot$, and a pair of M-dwarfs
of $\sim$0.5\,$M_\odot$ each on a tight orbit with a tentative period of
3.7\,d. The masses of A and Ba+Bb are nearly equal.

Further  spectroscopic  observations  of  $\kappa$~For are  needed  to
confirm the tentative  period of Ba,Bb and to  determine its orbit and
mass.   High-resolution spectroscopy  in the  near-infrared  where the
contrast  of  Ba,Bb is  more  favorable  will  help in  this  endeavor
\citep[e.g.][]{Bender2008}.   Given the short  period, eclipses in
Ba,Bb are likely, but their photometric detection presents a challenge
because of  the small (sub-percent) amplitude in  the combined visible
light.  Precise photometry will eventually detect flares on the active
M  dwarfs.  Long-term monitoring  of  both  orbits  will lead  to  the
accurate  measurement of  the distance  and  the masses  of all  three
stars.

Until  now,  the triple  nature  of  $\kappa$~For  was not  recognized
despite its  closeness to the  Sun.  The survey  of \citet{Raghavan10}
lists it only as a binary.  The question is how many more such systems
are we missing? This relates to the ways of discovering sub-systems in
the faint secondary companions.

The  binarity  of   B  was  revealed  by  its   unusually  high  mass.
Over-massive companions  can be  detected by astrometry  quite easily,
comparing the photo-center axis  $\alpha$ with the estimated full axis
$a$ (see eq.~\ref{eq:phi}), if  the astrometric orbit and parallax are
known.   However, reliable astrometric  orbits are  rare. Acceleration
measured  by {\it Hipparcos}  can also  be used  to estimate  the mass
ratio in  binaries with reliably  known visual or  astrometric orbits.
On the other hand, 
an  excess of the  total mass  in orbital  visual binaries  with known
parallax is a  less promising way of finding  such sub-systems because
of  the typically  large uncertainties  in both  measured  and modeled
masses.  Over-massive (binary) secondaries also can be detected by
a high minimum  mass inferred from a spectroscopic  orbit, as e.g.  in
the  case of  the quadruple  system  HD~27638 \citep{TG01,Torres2006}.
Over-massive   but  invisible  companions   could  be   white  dwarfs;
distinguishing them  from tight pairs  of red dwarfs  usually requires
some  sort  of  photometry.   When  the  resolved  photometry  of  the
secondary  companion in two  colors is  available (e.g.   from speckle
interferometry), it can be placed on  the CMD and detected as a binary
by an excess in luminosity.  Last but not least, spatial resolution of
the sub-systems is the  most direct and powerful method;  if the period of
Ba,Bb were  longer than $\sim$0.5\,yr (semi-major  axis $>30$\,mas), it
would have been resolved at SOAR.

Statistical modeling done  by \citet{NICI} hints that 10\%  to 20\% of
nearby solar-type astrometric binaries  found by {\it Hipparcos} could
have  ``massive'' secondaries  (sub-systems  or  white dwarfs).
This conclusion, however, depends on several assumptions in the model.
The detection techniques outlined  above can be applied systematically
to  nearby  binaries  to  better  constrain the  fraction  of  massive
secondaries  without making  any  assumptions. If  sub-systems in  the
secondary companions  are indeed as  frequent as in the  primaries, the
known number of hierarchical triples will be nearly doubled, changing
the multiplicity statistics dramatically.

Origins of  multiple stars are still actively  researched and debated,
and  different   theories  are  tested  by   their  predictions  about
multiplicity.  For example,  chaotic $N$-body dynamics rarely produces
pairs  of  low-mass stars  revolving  around  more massive  primaries;
rather, the primary  itself is likely to end up in  a close pair while
the  least massive  body is  ejected into  a distant  orbit  around it
\citep{DD03}.   Fragmentation  of rotating  cores  makes the  opposite
prediction where  the sub-systems  in primary or  secondary components
are  equally likely  to  form, being  a  natural sink  of the  angular
momentum.  In this  paradigm, stars are often born  as quadruples, but
some  inner sub-systems  subsequently merge.   Originally $\kappa$~For
could have consisted of two pairs of similar 0.5-$M_\odot$ stars.  One
pair  had merged  long time  ago and  became the  present-day primary,
while the other pair Ba,Bb is still here.

Binary  secondary companions  such  as the  one  of $\kappa$~For  have
implications for  the search  of exo-planets. If  the period  of Ba,Bb
were few months or years,  the weak secondary lines would always blend
with the lines  of A and create a small periodic  RV signal that could
be  mistaken  for  an   exo-planet  signature,  as  e.g.  in  HD~19994
\citep{Roell}.





\acknowledgments  I thank  M.~Giguere for  scheduling this  program at
CHIRON  and  running  the  data-reduction pipeline.   The  design  and
construction  of   CHIRON  was  supported  by   the  National  Science
Foundation under the ARRA AST-0923441.
This work  used the  SIMBAD service operated  by Centre  des Donn\'ees
Stellaires (Strasbourg, France)  and bibliographic references from the
Astrophysics Data System maintained by SAO/NASA.


%

{\it Facilities:}  \facility{SOAR},\facility{CTIO:1.5m}, \facility{Gemini-S}


\begin{thebibliography}{}

\bibitem[Abt \& Willmarth (2006)]{Abt06}
Abt, H. A., Willmarth, D. 2006, ApJS, 162, 207

\bibitem[Bertone et al. (2008)]{Bertone08}
Bertone, E., Buzzoni, A., Chavez, M., Rodriguez-Merino, L. H, 2008, A\&A, 
485, 823

\bibitem[Bender \& Simon (2008)]{Bender2008}
Bender, C. \& Simon, M. 2008, ApJ, 689, 416


\bibitem[Campbell \& Moore (1928)]{Lick}
Cambpell, W. W. \& Moore, J. H. 1928, Publ. Lick Obs. 16, 1

\bibitem[Chun et al. (2008)]{Chun08}
Chun, M. Toomey, D., Wahhaj, Z., Biller, B., 
Artigau, E., Hayward, T., Liu, M., Close, L., Hartung, M., Rigaut, F., \& Ftaclas, Ch.
2008, in: Adaptive Optics Systems. Ed. Hubin, N., Max, C. E., Wizinowich, P. L.
Proc. SPIE, 7015, 70151V



\bibitem[Delgado-Donate et al. (2003)]{DD03}
Delgado-Donate, E. J., Clarke, C. J., \& Bate, M.
2003, MNRAS, 342,  926

\bibitem[Endl et al. (2002)]{Endl02}
Endl M., K\"urster, M., Els, S. et al. 2002, A\&A, 392, 671


\bibitem[ESA (1997)]{HIP}
ESA 1997, The Hipparcos and Tycho Catalogues, ESA SP-1200


\bibitem[Girardi et al. (2000)]{Girardi}
Girardi, L., Bressan, A., Bertelli, G., \& Chiosi, C.
2000, A\&AS,141, 371

\bibitem[Gontcharov et al. (2000)]{Gontcharov2001}
Gontcharov, G. A., Andronova, A. A., Titov, O. A.
2000, A\&A, 355, 1164

\bibitem[Goncharov \& Kiyaeva (2002)]{GK02}
Goncharov, G. A., Kiyaeva, O. V. 2002, SvAL, 28, 261



\bibitem[Guedel et al. (1995)]{Guedel}
Guedel, M., Schmitt, J. H. M. M., Benz, A. O. 
1995, A\&A, 302, 775

\bibitem[Hartkopf et al. (2012)]{HTM12}
Hartkopf, W. I., Tokovinin, A., Mason, B. D., 2012, AJ, 143, 42




\bibitem[Hinkle et al. (2000)]{Arcturus}
Hinkle, K., Wallace, L., Valenti, J., \& Harmer, D. 2000,
Visible and Near Infrared Atlas of the Arcturus Spectrum
3727-9300\AA. 
(San Francisco: ASP) 

\bibitem[Lafreni\`ere et al. (2007)]{LAF2007}
Lafreni\`ere, D., Doyon, R.,  Marois, C., Nadeau, D., Oppenheimer, B.R.,
Roche, P.F., Rigaut, F., Graham, J.R., et al.
2007,  ApJ 670, 1367


\bibitem[Makarov \& Kaplan (2005)]{MK05}
Makarov, V. V. \& Kaplan, G. H.,  2005, AJ, 129, 2420

\bibitem[Mason et al. (2001)]{WDS}
Mason, B. D., Wycoff, G. L., Hartkopf, W. I., Douglass, G. G. \&
Worley, C. E. 2001, 
AJ 122, 3466 (see the current version at 
{\it http://www.usno.navy.mil/USNO/astrometry/optical-IR-prod/wds/wds.html})

\bibitem[Nakajima \& Morino (2012)]{Nakajima2012}
Nakajima, T. \& Morino, J.-I. 2012, AJ, 143, 2


\bibitem[Nidever et al. (2002)]{Nidever02}
Nidever, D. L., Marcy, G. W., Butler, R. P., Fischer, D. A., Vogt, S. S.
2002, ApJS, 141, 503


\bibitem[Nielsen \& Close (2010)]{Nielsen10}
Nielsen, E. L. \& Close, L. M. 2012, ApJ, 717, 878 


\bibitem[Nordstr\"om et al.  (2004)]{N04}
Nordstr\"om,  B., Mayor,  M.,  Andersen, J.,  Holmberg,  J., Pont,  F.,
Jorgensen, B. R., Olsen, E. H.,  Udry, S. \& Mowlavi, N.
2004, A\&A, 418, 989 



\bibitem[Raghavan et al. (2010)]{Raghavan10}
Raghavan, D., McAlister, H. A., Henry, T. J., Latham, D. W., 
Marcy, G. W., Mason, B. D., Gies, D. R., White, R. J., 
\& ten Brummelaar, Th. A. 2010, ApJS, 190, 1

\bibitem[R\"oll  et  al.  (2010)]{Roell}  R\"oll,  T.,  Seifahrt,  A.,
  Neuhäuser,  R.  \&  K\"ohler,  R.   2010,  in:  Binaries  ---  Key  to
  Comprehension  of the Universe.  Ed. Prsa,  A. \&  Zejda,  M. San
  Francisco: ASP Conf. Ser., V. 435


\bibitem[Schwab et al. (2012)]{CHIRON}
Schwab, Ch., Spronck, J., Tokovinin, A., Szymkowiak, A., Giguere, M.,
Fisher, D. 2012, Proc. SPIE, 8446,  paper 8446-9

\bibitem[Soubiran et al. (2010)]{PASTEL}
Soubiran, C., Le Campion, J.-F., Cayrel de Strobel, G., \& Caillo, A.
2010, A\&A, 515, A111


\bibitem[Tokovinin \& Gorynya (2001)]{TG01}
Tokovinin, A. A. \& Gorynya, N. A. 2001, A\&A, 374, 227



\bibitem[Tokovinin \& Cantarutti (2008)]{TC08}
Tokovinin, A., \& Cantarutti, R.  2008, PASP,  120, 170

\bibitem[Tokovinin et al. (2010)]{TMH10}
Tokovinin, A., Mason, B.D., Hartkopf, W.I. 2010, AJ, 139, 743


\bibitem[Tokovinin et al. (2012)]{NICI}
Tokovinin, A., Hartung, M., Hayward, Th. L., Makarov, V. V.  2012, AJ, 144, 7


\bibitem[Torres (2006)]{Torres2006}
Torres, G. 2006, AJ, 131, 1702


\bibitem[Trilling et al. (2008)]{Trilling08}
Trilling, D. E., Bryden, G., Beichman,  C. A. et al. 2008, ApJ, 674, 1086

\bibitem[van Leeuwen (2007)] {HIP2}
van Leeuwen, F. 2007, A\&A, 474, 653






\end{thebibliography}
\end{document}